\documentclass[english,keywords,showpacs,amsmath,amssymb,showpacs]{iopart}                                                       
\usepackage{graphicx}
\usepackage{color}
\usepackage{bm}
\usepackage{longtable}
\begin{document} 
\title{Quantitative probing of quantum-classical transition for the arrival time distribution$^{*}$}
\author{Dipankar Home,\footnote{dhome@bosemain.boseinst.ac.in}$^1$ Alok Kumar Pan\footnote{apan@bosemain.boseinst.ac.in}$^1$ and Arka Banerjee\footnote{arka.2110@gmail.com}$^2$}

\address{$^1$Department of Physics, Bose Institute, Calcutta 700009, India}

\address{$^2$St. Stephen's College, Delhi 11007, India}
\begin{abstract}
The classical limit problem of quantum mechanics is revisited on the basis of a scheme that enables a quantitative study of the way the quantum-classical agreement emerges while going through the intermediate mass range between the microscopic and the macroscopic domains. As a specific application of such a scheme, we investigate the classical limit of a quantum time distribution - an area of study that has remained largely unexplored. For this purpose, we focus on the arrival time distribution in order to examine the way the observable results pertaining to the quantum arrival time distribution which is defined in terms of the probability current density gradually approach the relevant classical statistical results  for an ensemble that corresponds to a Gaussian wave packet evolving in a linear potential. 
\vskip 8cm
{\large *Accepted for publication in \emph{J. Phys. A: Math. Theor.}}

\end{abstract}
\maketitle
\section{Introduction and the underlying basic scheme}
Over the years, the analysis of various aspects of the classical/macroscopic limit of quantum mechanics has attracted considerable attention[1-10]. Broadly speaking, there are two distinct strands of investigations related to the classical limit of quantum mechanics. One direction of study has been to delineate the way `classical-like behavior' can be obtained for any quantum mechanical micro-system under suitable conditions. The other line of study seeks to probe the macroscopic range of validity of quantum mechanics by examining as to what extent the quantum mechanical results in a suitably defined macroscopic regime agree with the corresponding results derived from classical mechanics. It is from the latter perspective that, in this paper, we investigate the quantum-classical correspondence for an observable \emph{time distribution} - an issue that has been hitherto neglected in the context of the classical limit aspect of quantum mechanics.

A particular aspect of the present paper is that the quantum-classical transition is probed in a \emph{quantitative way}.  For this purpose, in order to characterize the relevant macroscopic domain, we use `mass' as the \emph{parameter} that is varied to study the way the convergence of classical and the corresponding quantum results occurs by approaching the large mass values while going through the intermediate range. At this stage, some relevant remarks would be appropriate about the legitimacy of treating `mass' as a parameter, instead of taking it to be an operator. 

First, note that the scope of our analysis is restricted within the nonrelativistic domain being based on the Galilean invariant Schr${\ddot o}$dinger equation for the spin-0 particles. Now, given the Galilean invariance of the Schr${\ddot o}$dinger equation, one may recall an interesting theorem due to Bargmann\cite{bargmann} which states that, in nonrelativistic quantum mechanics, one cannot have a coherent superposition of states of different masses(for an elegant  proof of this theorem, using a suitable sequence of Galilean constant velocity transformations, see, for example, Kaempffer\cite{kaem}).  An insight into the physical justification for this theorem has been provided by Greenberger\cite{green} who showed that this restriction arises essentially because that in the  nonrelativistic domain, the `coordinate time' does not differ from the `proper time'(measured in the moving frame). One is, therefore, entitled to treat `mass' as a parameter, as long as the study is restricted within the framework of nonrelativistic quantum mechanics. 

Next, coming to the conceptual basis of the quantum-classical comparison scheme that  will be specifically used in this paper, we first note the following. While the predictions of the quantum mechanical formalism  are verifiable pertaining to essentially an ensemble of particles \cite{balleensemble, homewhitaker}, classical mechanics can describe the properties of an ensemble of particles as well as of a single particle.  Thus, the comparison between these two mechanics is operationally unambiguous provided one compares their statistical predictions for the dynamical evolutions of the \emph{same} given initial ensemble. 

It is in this spirit that we adopt the scheme used in this paper\cite{sengupta} where the quantum and the classical evolutions are compared by starting from the \textit{same} initial ensemble that has the specified position and momentum distributions obtained from a given wave function. While the quantum evolution is in accordance with the Schr${\ddot o}$dinger equation, the classical evolution of the given initial ensemble is calculated in terms of the classical phase space dynamics based on Liouville's equation.  However, a critical point in the classical calculation is the following. The initial phase space distribution for an ensemble is not uniquely fixed even if the position and the momentum distributions are specified. But, since in classical mechanics, the time evolution of all the usual observable properties of an ensemble are determined by the initial positions and momenta which are mutually independent variables, an initial phase space distribution ${D_0}(x_0,p_0,t=0)$  evolving under classical dynamics can be written, in the simplest possible choice, as a product of the position and momentum distributions pertaining to a given initial wave function $\Psi(x_0,t=0)$, given by

\begin{equation}
\label{dx0p0}
{D_0}(x_0,p_0,t=0)=|\Psi(x_0,0)|^2 \hskip 0.1cm |\Phi(p_0,0)|^2
\end{equation}
\noindent
where the variables $x_0$ and $p_0$ are the initial positions and momenta of the particles, and $\Phi(p_{0},0)$ is the Fourier transform of $\Psi(x_{0},0)$.

Based on this specific quantum-classical comparison scheme, the plan of this paper is as follows. In Section II we discuss the basic features of both the quantum and the classical procedures for defining the arrival time distribution using the probability current density. Here we may note that in recent years, the quantum mechanical distributions of various types of time like the tunneling time, arrival time, transit time, decay time, and so on have been widely studied; for useful reviews, see, for example, Muga \emph{et al.}\cite{mugabook} and Olkhovski\emph{ et al.}\cite{others}. In the light of this flourishing line of  works, the classical limit aspect of such quantum time distributions deserves to be a germane area of study. To this end, in this paper, we initiate such an investigation by restricting our attention to the classical limit of a particular form of quantum arrival time distribution that is defined in terms of the probability current density\cite{mugabook, muga92, ali03, pan06} - our analysis being contingent upon a specific scheme for the quantum-classical comparison,  and is couched in terms of a Gaussian wave packet propagating in a linear potential, while such a study, in principle, can be extended for other forms of time-distributions, using wave functions of various types, and in the context of any other potential.

In Section III, the quantum-classical correspondence of an arrival time distribution is treated in detail in terms of a general Gaussian wave packet (that does \textit{not} correspond to the minimum value of the uncertainty product $\Delta x \Delta p$) which evolves in the presence of a one dimensional linear potential. The salient feature of this work is a \emph{quantitative study} that is aimed at delineating the mass range over which the quantum results pertaining to the time distribution under consideration gradually concur with their classical counterpart - the representative relevant results for the mean arrival time and the associated fluctuation being given in section IV.  In the concluding Section V some future directions of work are indicated. 

But, before proceeding further, for the sake of completeness, some remarks are in order to stress the conceptual inadequacy of the usual textbook definition of the classical limit of quantum mechanics given in terms of $\hbar \rightarrow 0$. First, the notion that $\hbar \rightarrow$ is `small' has no absolute meaning because its value depends on the system of units\cite{berry91}. Further, wave functions are, in general, highly nonanalytic in the neighborhood of the limit point $\hbar \rightarrow 0$ \cite{berry72}. This results in the essential singularity of the quantum mechanically computed quantities at $\hbar \rightarrow 0$ . It is, thus, not possible to regard quantum mechanics as a perturbative extension of classical mechanics in the same sense as special relativity can be viewed as related to Newtonian mechanics by a convergent perturbation expansion in $v/c$\cite{berry89}. Hence, the only sensible operational formulation of the $\hbar \rightarrow 0$  classical limit condition would be to consider a dimensionless parameter of the form $\hbar/S << 1$ where $S$ is the `action quantity' relevant to a given situation. But, then, within this approach, an element of arbitrariness comes into play in the \textit{choice} of the appropriate `action quantity' to be used in any given example for probing the classical limit of quantum mechanics. In contrast, the procedure adopted in our paper for studying the macrolimit of quantum mechanics by varying `mass' as the relevant parameter is devoid of any such arbitrariness.

\section{Arrival time distributions in quantum and classical dynamics}
First, let us consider the 	quantum mechanical case. For simplicity, throughout this paper, we restrict the treatment to one spatial dimension. We begin with the non-relativistic quantum mechanical description of the flow of probability, expressed in terms of the position space distribution, that is governed by the continuity equation (derived from  the Schr${\ddot o}$dinger equation) given by
\begin{equation}
\label{con}
\frac{\partial}{\partial t}|\Psi({\bf x},t)|^2 + 
{\bf \nabla}.{J}({\bf x},t)=0
\end{equation}
The quantity ${\bf J}({\bf x},t)$=$\frac{i\hbar}{2m}(\Psi {\bf \nabla} 
\Psi^{\ast}-\Psi^{\ast}{\bf \nabla} \Psi)$, called  the probability current density, characterises this flow of probability. It is this current density that has been used in a number of studies to define the arrival time distribution for free particles\cite{mugabook,muga92,ali03}. By interpreting the equation of continuity in 
terms of the flow of physical probability, in conjunction with using the Born interpretation for the squared modulus of the wave function as denoting the probability density, it has been suggested that the mean quantum 
arrival 
time of the particles reaching a detector located at ${\bf x=X}$ may be 
written as
\begin{equation}
\label{qmeantime}
\left\langle t_Q\right\rangle =\frac{\int^{\infty}_{0}|{J}({\bf x}={\bf X}, t)| t  dt}
{\int^{\infty}_{0}|{J}({\bf x}={\bf X}, t)| dt}
\end{equation}
whence the corresponding fluctuation $(\Delta t)_{Q}$ is given by the root mean square deviation $(\Delta t)_{Q}=\sqrt{\langle t^{2}_{Q}\rangle- \langle t_{Q} \rangle^{2}}$. 

The definition of the mean arrival time specified by Eq.(\ref{qmeantime}) is, however, not a uniquely derivable result within standard quantum mechanics. In fact, different schemes for defining the quantum arrival time distribution have been discussed in the literature; for example, using Kijowski's axiomatic approach\cite{kijowski74}, or by invoking the time-of-arrival operator method in conjunction with the  POVM approach\cite{giannitrapani97}, by constructing  self-adjoint variants of the time-of-arrival operator\cite{grot96}, or by using the Bohmian causal model\cite{leavens90a}.  However, the ambit of the present paper is confined to only the probability current density approach\cite{mugabook, muga92,ali03}. Here it may be  noted that in certain situations, the quantity $J(X,t)$ can be negative during some time interval, even if the initial wave function has the positive momentum support - this is called the backflow effect \cite{bracken94}. It is in order to take this effect into account that the modulus of the quantity $J(X,t)$(suitably normalised) is taken for specifying $\langle t_{Q}\rangle$ as given by Eq.(\ref{qmeantime}).

Next, we note that the Schr${\ddot o}$dinger probability current defined in terms of the continuity equation has an inherent ambiguity. This is because the continuity equation remains satisfied with the addition of any divergence free term to the probability current. On this point, Holland \cite{holland99} has shown the uniqueness of the probability current for the spin-1/2 particles using the Dirac equation. On the other hand, for the spin-0 particles, using the Kemmer equation \cite{kemmer}, it has been demonstrated\cite{baere} that the non-relativistic limit of the Kemmer probability current is unique, whose expression turns out to be that of the  Schr${\ddot o}$dinger probability current. Hence, for the spin-0 particles, the Schr${\ddot o}$dinger probability current can be used for computing the arrival time distribution. Thus, even though the Schr${\ddot o}$dinger probability current is not directly observable, having no correspondence with an appropriate self-adjoint operator\cite{mugabook,wan},  it can have an observable manifestation for the spin-0 particles through the arrival time distribution. The latter is, in practice, a \textit{measurable quantity} - this point being underscored in various experimental contexts involving the  time-of-flight measurements\cite{schellekens05} concerning, for example, cold trapped atoms, with the quantum probability current being invoked in the relevant theoretical analysis\cite{ali06}. Besides, several theoretical models of `quantum clock'\cite{pan06,qclock} have been studied that bring out the empirical relevance of time distributions such as the arrival/transit time.  

Now, let us examine the classical procedure for computing the arrival time distribution. For this, a classical statistical ensemble of particles is described by the phase space density function $D(x,p,t)$ . Consequently, the classical position and momentum distribution functions are respectively $\rho_C (x,t)=\int D(x,p,t) dp $ and $\rho_C (p,t)=\int D(x,p,t) dx$, while $D(x,p,t)$ satisfies the classical Liouville equation given by 
\begin{equation}
\label{liou}
\frac{\partial D(x,p,t)}{\partial t}+ {\dot x} \frac{\partial D(x,p,t)}{\partial x}
+ {\dot p} \frac{\partial D(x,p,t)}{\partial p} =0
\end{equation}
Integrating the above equation with respect to $p$ one gets 
\begin{equation}
\label{lioucon}
\frac{\partial {\rho_C}(x,t)}{\partial t}+ \frac{\partial}{\partial x}
\left[\frac{1}{m} {\bar p}(x,t) {\rho_C}(x,t) \right]=0
\end{equation}
where ${\bar p}={\int p D(x,p,t) dp}/{\int D(x,p,t) dp}$ is the ensemble average of the momentum values of the individual particles.\\

Defining ${\bar v}(x,t)={{\bar p}(x,t)}/m$ as the ensemble average of the individual velocity values, we get
\begin{equation}
\label{lioucon1}
\frac{\partial {\rho_C}(x,t)}{\partial t}+ \frac{\partial}{\partial x}
{J_C}(x,t)=0
\end{equation}
where ${J_C}(x,t)= \rho_{C}(x,t){\bar v}(x,t)$. Thus, Eq.(\ref{lioucon1}) can be  regarded as  the equation of continuity characterising the classical time evolution of a statistical ensemble of particles. Using the expressions for $\rho_{C}(x,t)$ and ${\bar v}(x,t)$, the
classical probability 
current density  can then be written as 
\begin{equation}
\label{clcurr}
{J_C}(x,t)=\frac{1}{m} {\int p D(x,p,t) dp}
\end{equation}
Given this statistical description, the mean classical arrival time is given by 
\begin{equation}
\label{clmeantime}
\left\langle t_{C}\right\rangle =\frac{\int^{\infty}_{0}|{J_C}({\bf x}={\bf X}, t)| t  dt}
{\int^{\infty}_{0}|{J_C}({\bf x}={\bf X}, t)| dt}
\end{equation}
whence the corresponding fluctuation $(\Delta t)_{C}$ is given by the root mean square deviation $(\Delta t)_{C}=\sqrt{\langle t^{2}_{C}\rangle-\langle t_{C} \rangle ^{2}}$.

\section{Quantum-classical correspondence for a non-minimum-uncertainty-product wave packet propagating in a linear potential}
In this section, we compare the quantum and the classical results for the position, momentum and time distributions by considering a  general  non-minimum-uncertainty-product Gaussian wave packet propagating in a linear potential ($V=Kx$). 

Here we take the initial wave function $\Psi(x,0)$ and its Fourier transform $\Phi(p,0)$ to be given by
\begin{equation}
\label{inminp}
\Psi(x,0)=\frac{1}{(2 \pi {\sigma_0}^2)^{1/4} \sqrt{1+iC}} 
exp\left[-\frac{x^2}{4 {\sigma_0}^2(1+iC)}+ i k x\right]
\end{equation}

\begin{equation}
\label{inminm}
\Phi(p,0)=\left(\frac{2 {\sigma_0}^2}{\pi {\hbar}^2}\right)^{1/4} 
exp\left[- \frac{{\sigma_0}^2 (p-{\bar p})^2}{{\hbar}^2} (1+iC)\right]
\end{equation}
where the group velocity of the wave packet $u={\hbar k}/m={\bar p}/m$. 

Note that we have taken an initial Gaussian wave function $\Psi(x,0)$ which is {\it not} a minimum uncertainty state, i.e., $\Delta x \Delta p=(\hbar/2)\sqrt{1+C^2}$ $>$ $\hbar/2$, where $C$ is any real number - such a non-minimum-uncertainty-product state corresponds to what is known as a squeezed state \cite{robinett05}. In the presence of a linear potential, for such an initial wave function, the Schr${\ddot o}$dinger time evolved wave function $\Psi(x,t)$, and consequently  the probability current density  ${J_Q}(x,t)$ are respectively given by

\begin{eqnarray}
\fl
\label{tdminp}
\nonumber
\Psi(x,t)&=&\frac{1}{(2 \pi {\sigma_0}^2)^{1/4} \sqrt{1+i(C+\frac{\hbar t}
{2 m {\sigma_0}^2})}}\ exp\left[\frac{im}{\hbar}(u-\frac{K t}{m}) (x-\frac{u t}{2})-\frac {i {K}^2 {t}^3}{6 m \hbar}\right]\\
&&\times exp\left\{-\frac{( x- u t+\frac{1}{2}\frac{K}{m}{t}^2)^2}
{4 {\sigma_0}^2 \left[1+i(C+\frac{\hbar t}{2 m {\sigma_0}^2}) \right]}\right\}
\end{eqnarray}                                                                                
\begin{eqnarray}
\label{qmincurr}
{J_Q}(x,t)=\rho_Q(x,t)~\left\{u-\frac{K t}{m} +\frac{\hbar (C+\frac{\hbar t}{2 m {\sigma_0}^2}) (x-ut+\frac{1}{2}\frac{K}{m}{t}^2)}
{2 m {\sigma^{2}_{Q}(t)} } \right\}
\end{eqnarray}
where $\rho_Q(x,t)$ is the quantum mechanical position probability distribution function given by                                                                                                 
\begin{eqnarray}
\label{tdpp}
\rho_Q(x,t)=|\Psi(x,t)|^2=\frac{1}{\sqrt{2 \pi \sigma^{2}_{Q}(t)}} exp\left\{-\frac{( x- u t+\frac{1}{2}\frac{K}{m}{t}^2)^2}
{2 \sigma^{2}_{Q}(t)}\right\}
\end{eqnarray}
where $\sigma_{Q}(t)=\sigma_{0} \sqrt{1+(C+\frac{\hbar t}{2 m {\sigma_0}^2})^2 }$ is the width of a quantum wave packet that corresponds to the position probability distribution at any given instant $t$.

Next, we focus on calculating the probability current density $J(x,t)$ using the classical statistical evolution. For this purpose, in accordance with Eq.(\ref{dx0p0}), it is crucial that the initial phase space distribution $D_{0}(x_{0}, p_{0}, t=0)$ to be used for the classical calculations is fixed by the initial position and momentum distributions of the ensemble that are taken to be the {\it same} as the corresponding initial quantum distributions obtained from Eq.(\ref{inminp}) and (\ref{inminm}) respectively. Accordingly, the expression for $D_{0}(x_{0}, p_{0}, t=0)$ is given by
\begin{eqnarray}
\label{inminpm}
{D_0}(x_0,p_0,0)&=&|\Psi(x_0,0)|^2 \hskip 0.1cm |\Phi(p_0,0)|^2\\
 \nonumber
&=&\frac{1}{\pi \hbar \sqrt{1+C^2}} ~exp\left\{ -\frac{x_0^2}{2 {\sigma_0}^2 (1+C^2)}
-\frac{2 {\sigma_0}^2 (p_0 - {\bar p})^2}{{\hbar}^2} \right\}
\end{eqnarray}

Now, in order to obtain the time evolved classical phase space density function $D(x,p,t)$, we consider the classical Hamiltonian for the freely moving particles $H=p^2/2m+K x$, and Hamilton's equations  given by $x=p_0t/m -\frac{1}{2}\frac{K}{m}{t}^2+x_0$ and $p=p_{0}-K t$. Then one can write
$x_0=x-p_0t/m+\frac{1}{2}\frac{K}{m}{t}^2$ and $p_{0}=p+ K t$. By substituting these values of $x_0$ and $p_{0}$ in the
expression for $D_0(x_0,p_0,0)$ given by Eq.(\ref{inminpm}), we obtain the time evolved classical phase space distribution
function $D(x,p,t)$ given by
\begin{equation}
\label{tdminpm}
D(x,p,t)=\frac{1}{\pi \hbar \sqrt{1+C^2}}~exp\left\{ -\frac{(x-\frac{p t}{m}-\frac{1}{2}\frac{K}{m}{t}^2)^2}
{2 {\sigma_0}^2 (1+C^2)}-\frac{2 {\sigma_0}^2 (p+K t-{\bar p})^2}{{\hbar}^2} \right\}
\end{equation}

Now, substituting in Eq.(\ref{clcurr})the expression for the time evolved phase space distribution function $D(x,p,t)$ from Eq.(\ref{tdminpm}), the probability current density pertaining to this classical ensemble is given by 
\begin{equation}
\label{clmincurr}
{J_C}(x,t)=\rho_C (x,t)~ \left\{ u-\frac{K t}{m}+ \frac{(x-ut+\frac{1}{2}\frac{K}{m}{t}^2)\frac{\hbar^2 t}{2 m {\sigma_0}^2}}{2 m \sigma^{2}_{C}(t)} \right\} 
\end{equation}
whence the position probability distribution for this classical ensemble is given by
\begin{eqnarray}
\fl
\label{finalclpp}
\rho_C (x,t)=\int D(x,p,t) dp=\frac{1}{\sqrt{2 \pi \sigma^{2}_{C}(t)}}\times exp\left\{- \frac{(x-u t+\frac{1}{2}\frac{K}{m}{t}^2)^2}{2 \sigma^{2}_{C}(t)}\right\}
\end{eqnarray}
where $\sigma_{C}(t)=\sigma_{0} \sqrt{1+C^{2}+\frac{\hbar^{2} t^{2}}{4 m^{2} {\sigma_0}^{4}} }$ is the time-varying width of the classical position distribution function at an instant $t$. 

Note that this spreading of the statistical distribution  embodied in Eq.(\ref{finalclpp}) ensues from the classical Liouville evolution, and is, in general, different from the corresponding quantum spreading of a wave packet($\sigma_{Q}(t)\neq \sigma_{C}(t)$), unless $C=0$. This means that the quantum and the classical spreadings of the position distributions agree only if the initial Gaussian statistical distribution corresponds to the minimum-uncertainty-product, i.e., if initially, $\Delta x \Delta p=\hbar/2$. It, therefore, needs stressing that such a spreading is not essentially a quantum mechanical property of a propagating wave packet, but is a generic feature associated with a time-varying position probability distribution, quantum or classical. While   for the positive values of $C$, $\sigma_{Q}(t)$ is larger than $\sigma_{C}(t)$ for all times, for the negative values of $C$, $\sigma_{Q}(t)$ is always smaller that $\sigma_{C}(t)$.

On the other hand, Eqs.(\ref{qmincurr}) and (\ref{clmincurr}) clearly show  that the quantum and the classical probability currents are, in general, \textit{not} the same , i.e., $J_{Q}(x,t)\neq {J_C}(x,t)$. But, if one imposes the \emph{minimum-uncertainty-product condition} $C=0$, the quantum and the classical probability currents become the same, i.e., $J_{Q}(x,t)={J_C}(x,t)$. Also, interestingly, this condition $C=0$ ensures an exact agreement sans any limiting condition between  the quantum and the classical position probability distributions given by Eqs.(13) and (17) respectively. 

\section{Results of some relevant quantitative studies and their implications}
In order to make a systematic study of the way an agreement emerges between the quantum and the classical probability currents given by Eqs.(12) and (16) respectively, thereby leading to a matching of the corresponding mean arrival times and their fluctuations for the \emph{non-minimum-uncertainty-product} Gaussian wave function($C\neq 0$) under consideration, we proceed as follows. 
\begin{figure}[h]
{\rotatebox{0}{\resizebox{11.5cm}{4.0cm}{\includegraphics{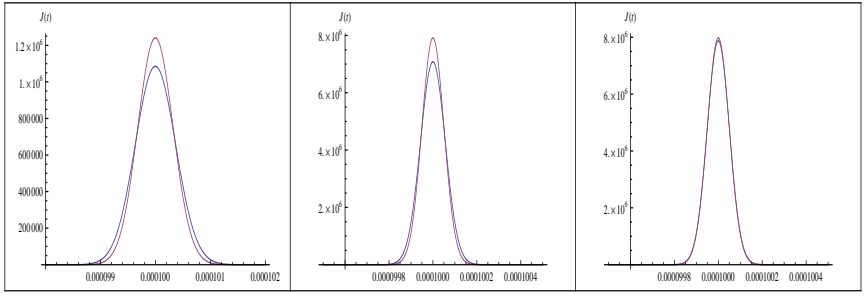}}}}
\caption{\label{fig} The time variations of the quantum and classical probability current densities $J_Q(x,t)$
and $J_C(x,t)$ are plotted for the values of m=1 a.m.u, m=50 a.m.u and m=720 a.m.u, corresponding to  ${\sigma_0}=10^{-5}$ cm, $u=10^4$ cm/sec, C=50 and $X=1$ cm. The bold and the dashed curves represent the quantum and classical cases respectively.}
\end{figure}
\begin{figure}[h]
{\rotatebox{0}{\resizebox{11.5cm}{4.0cm}{\includegraphics{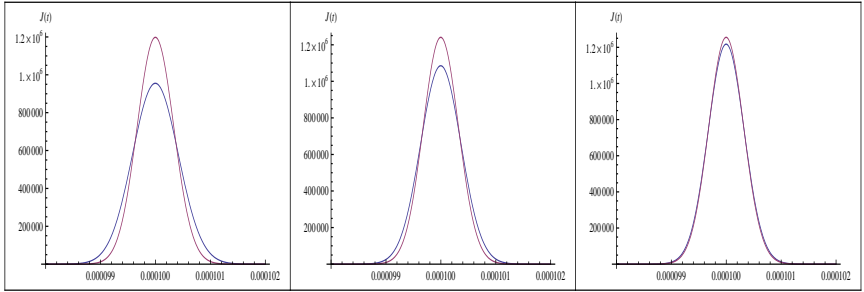}}}}
\caption{\label{fig} The time variations of the quantum and classical probability current densities  $J_Q(x,t)$
and $J_C(x,t)$ are plotted for the values of C=100, C=10 and C=1 by taking the  mass= 1 a.m.u corresponding to ${\sigma_0}=10^{-5}$ cm,
$u=10^4$ cm/sec, and X=$1$ cm. The bold and the dashed curves represent the quantum and classical cases respectively.}
\end{figure}
First, for a fixed value of $C$,  we compare the plots of the quantum and the classical probability currents by varying the values of the masses. As  representative studies, we take $C=50$ and the choices of the masses to be $m=1$ a.m.u(H atom), $m=50$ a.m.u and $m=720$ a.m.u($C_{60}$molecule). It is seen from Figure 1 that while for  $m=50$ a.m.u the disagreement between the quantum and the classical plots diminishes as compared to that for $m=1$ a.m.u, a complete agreement is ensured from the masses around $m=720$ a.m.u ($C_{60}$ molecule). In order to complement this line of study, another comparison is made between the plots of the quantum and the classical probability currents for a fixed mass, say, $m=1$ a.m.u by varying the values of C ranging over $C=100$, $C=10$ and $C=1$. Interestingly, it is seen from Figure 2 that while for $C=100$, the quantum and the classical curves appreciably differ, this difference gradually diminishes with the decreasing values of $C$(i.e., as the departure from the minimum-uncertainty-product Gaussian wave function gets minimised), with the difference becoming negligibly small as the value $C=1$ is reached.       
\begin{longtable}{|c|c|c|c|c|c|}
\hline 
$Mass(a.m.u)$& $\langle t_{C}\rangle$ (ms)& $\langle t_{Q}\rangle$ (ms)&$(\Delta t)_{C}$ (ms)&$(\Delta t)_{Q}$ (ms) \tabularnewline
\hline
\hline 
1&68.684&141.671&16.592&22.543\tabularnewline
5&13.008&14.575&8.069&9.523\tabularnewline
25&12.881&13.172&7.940&8.199\tabularnewline
50&12.877&13.025&7.936&8.064\tabularnewline
100&12.876&12.947&7.935&7.999\tabularnewline
500&12.876&12.890&7.935&7.947\tabularnewline
1000&12.876&12.883&7.935&7.941\tabularnewline
5000&12.876&12.877&7.935&7.936\tabularnewline
10000&12.876&12.876&7.935&7.935\tabularnewline
\hline
\end{longtable} 
{\footnotesize Table 1. The comparisons between quantum and classical mean arrival times and their respective fluctuations are given for the different values of mass, corresponding to a fixed value of $C=1000$, and the other relevant parameters being $u=10 cm/s$, $X=0.1cm$, $\sigma_{0}=10^{-4} cm$.}

Now, we come to a crucial aspect of this quantitative study; i.e., the probing of the range of masses over which an agreement emerges between the quantum and the classical  mean arrival times, as well as between their respective fluctuations. For this we proceed as follows. We take three different values  of $C$, viz. $C=1000$, $C=500$ and $C=100$, and, for any such given value of $C$, we vary the masses ranging from 1 a.m.u(H atom) to the heavier  molecules, say, biomolecules with molecular weights around $10^{3} - 10^{4} a.m.u$ (i.e., biomolecules comprising approximately 10-300 base pairs of DNA molecules, where 1 base pair $\approx$ 650 a.m.u). 

\begin{longtable}{|c|c|c|c|c|c|}
\hline 
$Mass(a.m.u)$& $\langle t_{C}\rangle$ (ms)& $\langle t_{Q}\rangle$ (ms)&$(\Delta t)_{C}$ (ms)&$(\Delta t)_{Q}$ (ms) \tabularnewline
\hline
\hline 
1&115.021&150.207&22.686&24.223\tabularnewline
5&10.397&11.324&4.808&5.654\tabularnewline
25&10.281&10.456&4.711&4.865\tabularnewline
50&10.277&10.365&4.708&4.784\tabularnewline
100&10.276&10.321&4.707&4.745\tabularnewline
500&10.276&10.285&4.707&4.715\tabularnewline
1000&10.276&10.281&4.707&4.711\tabularnewline
5000&10.276&10.276&4.707&4.707\tabularnewline
\hline
\end{longtable} 
{\footnotesize Table 2. The comparisons between quantum and classical mean arrival times and their respective fluctuations are given for the different values of mass, corresponding to a fixed value of $C=500$, and the other relevant parameters being $u=10 cm/s$, $X=0.1cm$, $\sigma_{0}=10^{-4} cm$.}

Then, from the relevant computational results as given in Table 1, corresponding to $C=1000$, it is seen that  an  appreciable difference between the quantum and the classical mean arrival times, as well as a significant difference between their respective fluctuations persist up to masses around $10^3$ a.m.u, after which  these differences gradually diminish. Eventually, these differences disappear beyond the mass range of $10^4$ a.m.u(say, for the protein molecule such as cytochrome-c having the mass $12\times 10^3$ a.m.u). It may also be noted that while the variations of both the quantities $\langle t \rangle_{C}$ and $(\Delta t)_{C}$ as the mass changes saturate at the mass value of $10^2$ a.m.u, the corresponding variations of both the quantities $\langle t \rangle_{Q}$ and $(\Delta t)_{Q}$ with mass saturate around the mass value $10^4$ a.m.u.

\begin{longtable}{|c|c|c|c|c|c|}
\hline 
$Mass(a.m.u)$& $\langle t_{C}\rangle$ (ms)& $\langle t_{Q}\rangle$ (ms)&$(\Delta t)_{C}$ (ms)&$(\Delta t)_{Q}$ (ms) \tabularnewline
\hline
\hline 
1&201.172&187.203&28.556&27.540\tabularnewline
5&10.124&10.321&1.206&1.697\tabularnewline
25&10.037&10.126&1.064&1.329\tabularnewline
50&10.001&10.002&1.003&1.128\tabularnewline
100&10.000&10.009&1.000&1.032\tabularnewline
500&10.000&10.002&1.000&1.006\tabularnewline
1000&10.000&10.000&1.000&1.000\tabularnewline
\hline
\end{longtable}
{\footnotesize Table 3. The comparisons between quantum and classical mean arrival times and their respective fluctuations are given for the different values of mass, corresponding to a fixed value of $C=100$, and the other relevant parameters being $u=10 cm/s$, $X=0.1cm$, $\sigma_{0}=10^{-4} cm$.}\\ 

The results similar to that given in Table 1 are presented in Table 2 and Table 3 for $C=500$ and $C=100$ respectively. While for $C=500$, the agreement between $\langle t \rangle_{C}$ and $\langle t \rangle_{Q}$, as well as between $(\Delta t)_{C}$and $(\Delta t)_{Q}$ emerge from the value of m=5000 a.m.u(say, for the insulin molecule having m=5808 a.m.u), for $C=100$, such an agreement is ensured from m=1000 a.m.u(say, for the peptide hormone molecule oxytocin having m=1007 a.m.u).

It is, therefore, seen from these numerical computations that for a given set of values of the parameters $u$, $X$, and $\sigma_{0}$, the greater the value of $C$ signifying an increasing deviation from the minimum-uncertainty-product Gaussian wave packet, the larger is the value of mass from which the quantum-classical agreement occurs for both the mean arrival time and its fluctuation. In other words, for the type of example studied here, by increasing the values of the parameter $C$, one can extend the range of mass values covering the heavier molecules for which appreciable  disagreements can be found  between the quantum and the classical values of the mean arrival time and its fluctuation. Thus, as regards the possibility of the relevant empirical studies, it would be interesting to probe the experimental realizability of the Gaussian wave packets that are characterised by large values of the parameter $C$. 
\section{Concluding remarks}
An application of the specific quantum-classical comparison scheme underlying this paper is currently under consideration by using such a scheme for analysing the classical limit of quantum time distributions which are appropriately defined in the context of the harmonic oscillator potential - this study intends to use both the minimum as well as the non-minimum-uncertainty-product initial wave packets, including the case of the initial minimum-uncertainty-product wave packet having a specific width that corresponds to what is known as Schr${\ddot o}$dinger's coherent state. Besides, further investigations are required to be pursued  along, say,  the following directions:

$a)$ The treatment presented in  this paper uses essentially  a Gaussian wave packet. It should, therefore, be interesting to study in terms of a suitably constructed non-Gaussian wave packet, or by using a superposition of wave packets, the extent to which the calculated quantitative results are dependent on  the form of the wave packet  being strictly Gaussian. Such a study, also using forms of potential other than the linear one, would be particularly helpful for examining the empirical feasibility of tests related to the type of example discussed in this paper.

$b)$ Since the initial phase space distribution function used in our classical calculation is not uniquely fixed even if the position and momentum distribution functions are specified for a given wave function, it would be instructive to compare the quantitative results of this paper with the corresponding results for the choices of the initial classical phase space distribution function other than the simplest possible choice we have used for the given position and momentum distributions. As a special case of such a choice, for a given non-minimum-uncertainty-product Gaussian wave function, one may adopt the prescription given by Wigner\cite{wigner} for fixing the initial phase space distribution function to be used in our calculation. Studies along this line, based on the specific quantum-classical comparison scheme that has been invoked in the present paper, should be useful in throwing light on the extent to which the delineation of the mass values over which the quantum-classical agreement emerges for the mean arrival time and its fluctuation is sensitive to the choice of the initial classical phase space distribution .

$c)$ The quantum arrival time distribution used in this paper has been calculated specifically in terms of the probability current density. Comprehensive investigations are required in order to compare the quantitative results given in this paper with those obtained from various other schemes[17-21, 25-28] that have been suggested for defining the quantum arrival time distribution. This line of study would have an added ramification as regards the important issue concerning the possibility of discriminating between the different quantum approaches  that have been proposed to compute the  arrival time distribution.  

$d)$ We note that the time-of-flight image method\cite{schellekens05} has been widely employed for inferring the temperature of a cloud of trapped atoms in the experiments which involve the laser cooling of atoms. In this context, a quantitative study of the way the semi-classically computed time-of-flight distributions match with the corresponding quantum distributions in the limits of large mass and high temperature is of considerable interest\cite{ali06}. It should, therefore, be relevant to take a fresh look at this issue by invoking the specific scheme for quantum-classical comparison that has been used in this paper. 

\section*{Acknowledgments}
DH is indebted to late Shyamal Sengupta for the insightful interactions that motivated this paper. AKP acknowledges helpful discussions related to this work during his visits to the Perimeter Institute, Canada; Centre for Quantum Technologies, National University of Singapore; and Benasque Centre for Science, Spain. DH thanks DST, Govt. of India, and the Centre for Science and Consciousness, Kolkata for support. AKP acknowledges the Research Associateship of Bose Institute, Kolkata. 

\end{document}